\documentclass[letterpaper, 10 pt,  conference]{ieeeconf}  

\usepackage{graphicx} 
\usepackage{amsmath,bm,times} 
\usepackage{amssymb}  
\usepackage{tikz}
\usetikzlibrary{shapes,arrows,backgrounds,fit,positioning}
\usepackage{subfigure}
\usepackage{cite}
\usepackage{soul}
\usepackage{balance}
\usepackage{tabularx}
\usepackage{url}

\newtheorem{assumption}{Assumption}
\newtheorem{problem}{Problem}

\newtheorem{remark}{Remark}
\newtheorem{theorem}{Theorem}
\newtheorem{definition}{Definition}

\newtheorem{lemma}{Lemma}
\newtheorem{example}{Example}

\renewcommand{\theenumi}{(C\arabic{enumi}}

\title{\LARGE \bf Tuning Rate of Strategy Revision in Population Games}%
\author{Shinkyu Park 
  \thanks{This work was supported by funding from King Abdullah University of Science and Technology (KAUST).}
  \thanks{Park is with the Electrical and Computer Engineering, King Abdullah University of Science and Technology (KAUST), Thuwal 23955, Saudi Arabia. {shinkyu.park@kaust.edu.sa}}
}

\IEEEoverridecommandlockouts
\begin{document}

\maketitle

\begin{abstract}
  We investigate a multi-agent decision problem in population games where each agent in  a population makes a decision on strategy selection and revision to engage in repeated games with others. 
  The strategy revision is subject to time delays which 
  represent the time it takes for an agent revising its strategy needs to spend before it can adopt a new strategy and return back to the game. We discuss the effect of the time delays on long-term behavior of the agents' strategy revision. In particular, when the time delays are large, the strategy revision would exhibit oscillation and the agents spend substantial time in ``transitioning'' between different strategies, which prevents the agents from attaining the Nash equilibrium of the game. As a main contribution of the paper, we propose an algorithm that tunes the rate of the agents' strategy revision and show such tuning approach
  ensures convergence to the Nash equilibrium. We validate our analytical results using simulations.
\end{abstract}

\section{Introduction}
Consider a multi-agent decision problem where each agent selects a strategy to engage in repeated interactions with other agents. Agents receive payoffs as a function of their \textit{strategy profile} -- the distribution of their strategy selection -- and revise their strategy selection based on the payoffs. Such problem setting is prevalent in many engineering applications \cite{8647522, 7244342, 7379949, 5345810, 7422154, doi:10.1080/00207179.2016.1231422, 7823106, 5161293, 10.2307/25768135, 9561809}  ranging from demand response in smart grids to task allocation in multi-robot systems research.

We adopt the population game formalism \cite{Sandholm2010-SANPGA-2} to model the process of the agents' strategy revision and assess long-term behavior of the revision process. Different from conventional population game problems, in this work, we investigate the scenario where the strategy revision is subject to time delays for which every agent revising its strategy needs to spend a fixed amount of time for transitioning to a new strategy. As a case in point, in multi-robot task allocation applications \cite{5161293, doi:10.1177/0278364911401442, 9561809} where the strategy revision corresponds to assignment of a new task to a robot, time delays in the strategy revision reflect the requirement for each robot to travel to a distant location to take on a new task.

When the agents' strategy revision is subject to time delays, as we discuss in the paper, their strategy profile would exhibit oscillation implying that a portion of the agent population persistently transitions between different strategies. In the task allocation applications, this means that the robots would spend substantial time on traveling between distant locations which is a disadvantage since during the traveling, they cannot carry out assigned tasks. The authors of \cite{5161293, doi:10.1177/0278364911401442} concretely discuss such phenomenon observed in their linear models, and empirical results presented in \cite{9561809} illustrate the oscillation of the strategy profile in population game models.

As a main contribution, we propose an algorithm that tunes the rate of the agents' strategy revision to eliminate the oscillation of the strategy profile, and prove that the algorithm ensures convergence of the strategy profile to the Nash equilibrium. Of relevance to this work are \cite{9483414},\cite[Section~5.9]{fox_phdthesis} and references therein that investigate the scenarios where payoff mechanisms underlying population games are subject to time delays and hence the agents revise their strategy selection based on time-delayed payoffs. Also, \cite{9993228} examines a relevant problem in which, when revising its strategy, each agent needs to take on a sequence of sub-tasks (sub-strategies) which would potentially cause time delays in the agent's strategy revision.

Our contributions are distinct from existing work in the following aspects: \textbf{(i)} unlike \cite{9483414},\cite[Section~5.9]{fox_phdthesis} and references therein, our problem formulation considers that the agents' strategy revision is subject to time delays, which is different from the delays in payoff mechanisms, 
and \textbf{(ii)}  different from \cite{5161293, doi:10.1177/0278364911401442, 9993228}, our proposed approach ensures convergence of the strategy profile to the Nash equilibrium, which, in our problem formulation, requires that no agents are in transition or taking on any sub-strategies. 
Below we summarize the main results of the paper.
\begin{itemize}
\item We model and analyze the effect of time delays in the strategy revision using the population game formalism and discuss long-term behavior of the agents' strategy profile. In particular, our analysis explains how the rate of the agents' strategy revision affects the convergence of the strategy profile.
  
\item Based on the analysis, we propose an algorithm for tuning the strategy revision rate. The algorithm judiciously decreases the revision rate and ensures the convergence of the strategy profile to the Nash equilibrium in \textit{contractive} population games. We illustrate our analytical results using simulations.
  
\end{itemize}

The paper is organized as follows. 
In Section~\ref{section:problem_description}, we introduce the population game framework and strategy revision model we adopt in this study. 
We also illustrate how the rate of the agents' strategy revision affects long-term behavior of their strategy profile. In Section~\ref{section:revision_rate_scheduling}, we propose an algorithm that judiciously tunes the strategy revision rate and 
show that the proposed algorithm ensures the convergence of the strategy profile to the Nash equilibrium in the class of \textit{contractive} population games.\footnote{See Definition~\ref{definition:contractive_games} for its formal definition.}
We illustrate our analytical results using simulations with a numerical example in Section~\ref{section:simulations}. We conclude the paper with discussions and future plans in Section~\ref{section:conclusions}.

\paragraph*{Notation} We denote by $\mathbb R^n$ the set of $n$-dimensional real vectors and by $\mathbb R_+^n$ the set of (element-wise) non-negative $n$-dimensional real vectors. Throughout the paper, we adopt the Euclidean norm for matrices and vectors.

\section{Problem Description} \label{section:problem_description}
We describe the foundation of the population game formalism \cite{Sandholm2010-SANPGA-2} and explain how we adopt it in our multi-agent decision problem to study the effect of time delays in strategy revision.

\subsection{Population Games and Evolutionary Dynamics} \label{section:population_games}
Consider that a population of decision-making agents are engaged in repeated strategic interactions with one another. Each agent has a finite set $\{1, \cdots, n\}$ of strategies available and can select one strategy at a time. Based on its strategy selection, the agent receives a payoff determined by a payoff mechanism underlying the interactions.

Adopting the conventional notation and formalism, we denote the state of the population by $x = (x_1, \cdots, x_n) \in \mathbb R_+^n$, where $x_i$ represents the portion of the population selecting strategy~$i$ and $x$ sums up to $1$, i.e., $\sum_{i=1}^n x_i = 1$. A payoff vector $p = (p_1, \cdots, p_n) \in \mathbb R^n$ is determined by a continuous function $\mathcal F$ as $p = \mathcal F(x)$, where $p_i$ is the payoff assigned to the agents selecting strategy~$i$.
Assuming that there are a large number of agents in the population, we can define the set of viable population states as $\mathbb X \!=\! \{ z \in \mathbb R_+^n | \sum_{i=1}^n z_i \!=\! 1 \}$. 

Contractive population games are defined as follows.
\begin{definition} [Contractive Population Game \cite{SANDHOLM2015703}] \label{definition:contractive_games}
  A population game $\mathcal F$ is called \textit{contractive} if it holds that
  \begin{align} \label{eq:contractive_game}
    (w - z)^T (\mathcal F(w) - \mathcal F(z)) \leq 0, ~ \forall w,z \in \mathbb X.
  \end{align}
\end{definition}

\begin{assumption} \label{assumption:F_bounded}
  We assume that the differential map $D\mathcal F$ of $\mathcal F$ exists and is bounded: there is a positive constant $B_{D\mathcal F}$ satisfying $\|D\mathcal F(z)\|_2 \leq B_{D\mathcal F}, ~ \forall z \in \mathbb X$.
\end{assumption}

Note that when $D\mathcal F$ exists, the requirement \eqref{eq:contractive_game} can be cast as
\begin{align} \label{eq:contrative_game_DF}
  \tilde z^T D\mathcal F(z) \tilde z \leq 0, ~\forall z \in \mathbb X, \tilde z \in T\mathbb X,
\end{align}
where $T\mathbb X$ is the tangent space of $\mathbb X$.

The Nash equilibrium of $\mathcal F$ is defined as follows.
\begin{definition} [Nash Equilibrium] \label{definition:NE}
  An element $z^{\footnotesize \text{NE}}$ in $\mathbb X$ is called the \textit{Nash equilibrium} of a population game $\mathcal F$ if it holds that
  \begin{align}
    ( z^{\footnotesize \text{NE}} - z )^T \mathcal F(z^{\footnotesize \text{NE}}) \geq 0, ~ \forall z \in \mathbb X.
  \end{align}
\end{definition}
A population game $\mathcal F$ can have multiple Nash equilibria. We denote by $\mathbb{NE}(\mathcal F)$ the set of Nash equilibria. We adopt the following example to illustrate our main results.\footnote{For a simple presentation of the paper, we use the RPS game for the illustration purpose.}


\begin{example} \label{example:rps_game}
  Consider the Rock-Paper-Scissors (RPS) game ($n=3$) whose payoff function is given by
  \begin{align} \label{eq:payoff_function_rps}
    \mathcal F(x) =
    \begin{pmatrix}
      -a x_2 + b x_3 \\
      b x_1 - a x_3 \\
      -a x_1 + b x_2
    \end{pmatrix},
  \end{align}
  where $a,b$ are positive constants. Note that \eqref{eq:payoff_function_rps} is a contractive game if $b \geq a$, and the Nash equilibrium of $\mathcal F$ is given by $(1/3, 1/3, 1/3)$ for any $a,b > 0$.
\end{example}


In our study, we consider that the agents are repeatedly engaged in a same game and given an opportunity, they can revise their strategy selection based on the Smith revision protocol, originally presented in \cite{10.2307/25768135} and defined as follows.\footnote{Although the  analysis we present in this paper can be extended to other class of strategy revision protocols. However, for a concise presentation of our main result, we adopt the Smith protocol.}
\begin{align} \label{eq:smith_revision_protocol}
  \rho_{ij} (p)
  = \varrho [p_j - p_i]_+
  := \begin{cases}
    \varrho(p_j - p_i) & \text{if } p_j \geq p_i \\
    0 & \text{otherwise}
  \end{cases},
\end{align}
where 
$\varrho$ is a positive constant satisfying\footnote{Such constant $\varrho$ exists since $\mathcal F$ is a continuous function and $\mathbb X$ is a compact subset of $\mathbb R^n$.}
\begin{align} \label{eq:revision_protocol_inequality}
  \textstyle\sum_{j=1}^n \varrho [p_j - p_i]_+ \leq 1 \text{~~and~~} \varrho [p_j - p_i]_+ \leq 1.
\end{align}
According to \eqref{eq:smith_revision_protocol}, each agent switches its strategy selection $i$ to $j$ with probability $\varrho [p_j - p_i]_+$; otherwise, it stays with its current strategy selection~$i$ with probability $1 - \sum_{j=1}^n \varrho [p_j - p_i]_+$. Hence, higher the payoff associated with strategy $j$ more likely the agent selects it. The condition \eqref{eq:revision_protocol_inequality} is required for such probabilistic strategy revision scheme to be well-defined.

The point in time at which the agents' strategy revision takes place is determined by identical and independent Poisson processes with parameter $\lambda$. In particular, each agent can revise its strategy selection at each jump time of an associated Poisson  process. Since the parameter $\lambda$ determines the rate of the strategy revision, we refer to it as the \textit{strategy revision rate}.

Let $x(t) = (x_1(t), \cdots, x_n(t))$ be the population state at time instant $t$ driven by the revision protocol \eqref{eq:smith_revision_protocol}, the Poisson processes, and payoffs determined by a population game $\mathcal F$.
By same discussions presented in \cite[Chapter~10]{Sandholm2010-SANPGA-2}, as the population size tends to infinity, the solution to the following state equation approximates the population state $x(t)$ with arbitrary accuracy.
\begin{align} \label{eq:original_edm}
  \dot x_{i} (t)
  &= \mathcal V_i (x(t), p(t)) \nonumber \\
  &:= \lambda \big( \textstyle\sum_{j=1}^n x_j(t) \varrho \left[ p_i(t) - p_j(t) \right]_+ \nonumber \\
  &\qquad \qquad \qquad - x_i(t) \textstyle\sum_{j=1}^n \varrho \left[ p_j(t) - p_i(t) \right]_+ \big)
\end{align}
Adopting the same naming convention as in \cite{9029756}, we refer to \eqref{eq:original_edm} as \textit{Evolutionary Dynamics Model (EDM)}.

\subsection{$\delta$-Passivity Tools for Convergence Analysis}
A focal research theme in population games is in establishing convergence of the population state $x(t)$, governed by \eqref{eq:original_edm}, to the Nash equilibrium set $\mathbb{NE}(\mathcal F)$ of an underlying population game $\mathcal F$. An earlier work \cite{HOFBAUER20091665} in economics literature uses Lyapunov stability theorems to establish the convergence for a class of EDMs in contractive population games. Later, the work of \cite{Fox2013Population-Game} introduces the notion of $\delta$-passivity, originated from dynamical system theory \cite{willems_dissipative_1972}, in population games and provides general tools for studying convergence properties of EDMs. Since then, there have been refined definitions and applications of $\delta$-passivity proposed 
for convergence analysis in population games \cite{9029756, Park2019Payoff-Dynamic-, 9219202}.
Aside from $\delta$-passivity tools relevant to our work, there are other important works that present a different notion of passivity as a tool to study convergence in 
multi-agent games \cite{9022871, 9781277}.

We adopt the following definition of $\delta$-passivity for \eqref{eq:original_edm} from \cite{9029756}.
\begin{definition}
  EDM \eqref{eq:original_edm} is \textit{$\delta$-passive} if there is a continuously differentiable function $\mathcal S: \mathbb X \times \mathbb R^n \to \mathbb R_+$ for which
  \begin{multline}
    \mathcal S(x(t), p(t)) - \mathcal S(x(t_0), p(t_0)) \leq \\
    \int_{t_0}^t \dot x^T(\tau) \dot p(\tau) \, \mathrm d\tau, ~ \forall t \geq t_0 \geq 0
  \end{multline}
  holds for every payoff vector trajectory $p(t), ~ t \geq 0$. We refer to $\mathcal S$ as the \textit{$\delta$-storage function}.
\end{definition}

According to \cite{Fox2013Population-Game, 9029756}, EDM \eqref{eq:original_edm}, which is constructed using the Smith revision protocol \eqref{eq:smith_revision_protocol}, is $\delta$-passive and its $\delta$-storage function is given by
\begin{align} \label{eq:smith_storage_function}
  \mathcal S(x, p) = \frac{\lambda}{2} \textstyle\sum_{i=1}^n \textstyle\sum_{j=1}^n x_i \varrho [p_j - p_i]_+^2.
\end{align}
We make the following two observations which are used to establish the main result of the paper: for all $i,j$ in $\{1, \cdots, n\}$, it holds that
\begin{align*}
  \lambda^{-1} \mathcal S(x, p) \geq \frac{1}{2} x_i \varrho[p_j - p_i]_+^2 \geq \frac{1}{2} x_i^2 \varrho[p_j - p_i]_+^2 \geq 0
\end{align*}
from which we conclude that
\begin{align} \label{eq:S_convergence}
  \lambda^{-1} \mathcal S(x, p) \to 0 \implies x_i \varrho[p_j - p_i]_+ \to 0.
\end{align}
Also, by similar arguments used in the proof of Theorem~4.5 in \cite{Fox2013Population-Game}, we can verify that 
\begin{align} \label{eq:derivative_S_negative}
  \nabla_x^T \mathcal S(x,p) \mathcal V(x,p) \leq 0, ~ \forall x \in \mathbb X, p \in \mathbb R^n.
\end{align}

\subsection{Analyzing the Effect of Time Delays in Strategy Revision} \label{sec:analysis_transition_delay}
We consider that the agents' strategy revision is subject to  time delays: \textbf{(i)} an agent revising its strategy is required to spend a constant time period $d_{ji}$ to conclude its strategy revision from $j$ to $i$, and \textbf{(ii)} during the strategy revision, the agent cannot participate in the game. As a case in point, in the task allocation applications, the robots that are in transition to distant locations to take on newly assigned tasks cannot carry out the tasks until they arrive at the new locations.
To model this, we interpret $x_i(t)$ as the portion of the population that are engaged in the game with strategy selection $i$ at time~$t$, which \textit{exclude} the agents transitioning to strategy~$i$. Accordingly, the payoff vector $p(t) = \mathcal F(x(t))$ depends only on the agents who are participating in the game.

Note that in this problem setting, the population state $x(t)$ may not sum up to $1$, i.e., $\sum_{i=1}^n x(t) \leq 1$. For this reason, we extend the definition of the population state space as 
\begin{align}
  \mathbb X^{\footnotesize \text{ext}} = \left\{ z \in \mathbb R_+^n \,\big|\, \textstyle \sum_{i=1}^n z_i \leq 1 \right\}.
\end{align}
Accordingly, we extend the domains of the payoff function $\mathcal F: \mathbb X^{\footnotesize \text{ext}} \to \mathbb R^n$, the vector field $\mathcal V_i: \mathbb X^{\footnotesize \text{ext}} \times \mathbb R^n \to \mathbb R$, and the $\delta$-storage function $\mathcal S: \mathbb X^{\footnotesize \text{ext}} \times \mathbb R^n \to \mathbb R_+$. Also, we restate the boundedness condition of the map $D \mathcal F$ in Assumption~\ref{assumption:F_bounded} as $\max_{z \in \mathbb X^{\text{ext}}} \| D \mathcal F (z) \|_2 \leq B_{D \mathcal F}$.

Denote by $y_{ji}(t)$ the portion of the population that are in transition for the strategy revision from $j$ to $i$ at time~$t$. Given a fixed strategy revision rate $\lambda$, the following state equation describes how $y_{ji}(t)$ changes over time.
\begin{subequations}  \label{eq:y_ij_dynamics}
\begin{align}
  &\dot y_{ji} (t) = \lambda x_j(t) \varrho \left[ p_i(t) - p_j(t) \right]_+, ~ d_{ji} > t \geq 0 \\
  &\dot y_{ji} (t) = \lambda \big( x_j(t) \varrho \left[ p_i(t) - p_j(t) \right]_+ \nonumber \\
  &-\!x_j(t \!-\! d_{ji}) \varrho \left[ p_i(t \!-\! d_{ji}) \!-\! p_j(t \!-\! d_{ji}) \right]_+ \big), ~t \geq d_{ji} \label{eq:y_ij_dynamics_b}
\end{align}
\end{subequations}
with $y_{ji}(0) = 0$. Note that in \eqref{eq:y_ij_dynamics_b}, the first term $x_j(t) \varrho [ p_i(t) \!-\! p_j(t) ]_+$ denotes the rate at which the agents decide to revise their strategy selection from $j$ to $i$ and initiates transition from $j$ to $i$ at time~$t$. Whereas the second term $x_j(t \!-\! d_{ji}) \varrho \left[ p_i(t \!-\! d_{ji}) \!-\! p_j(t \!-\! d_{ji}) \right]_+$ describes the rate at which the agents complete the transition and return back to the game after spending the time period $d_{ji}$ in the transition.

\begin{figure}
  \center
  \subfigure[Population state $x(t)$]{
    \includegraphics[trim={.1in .1in .1in 0in}, clip, width=1.55in]{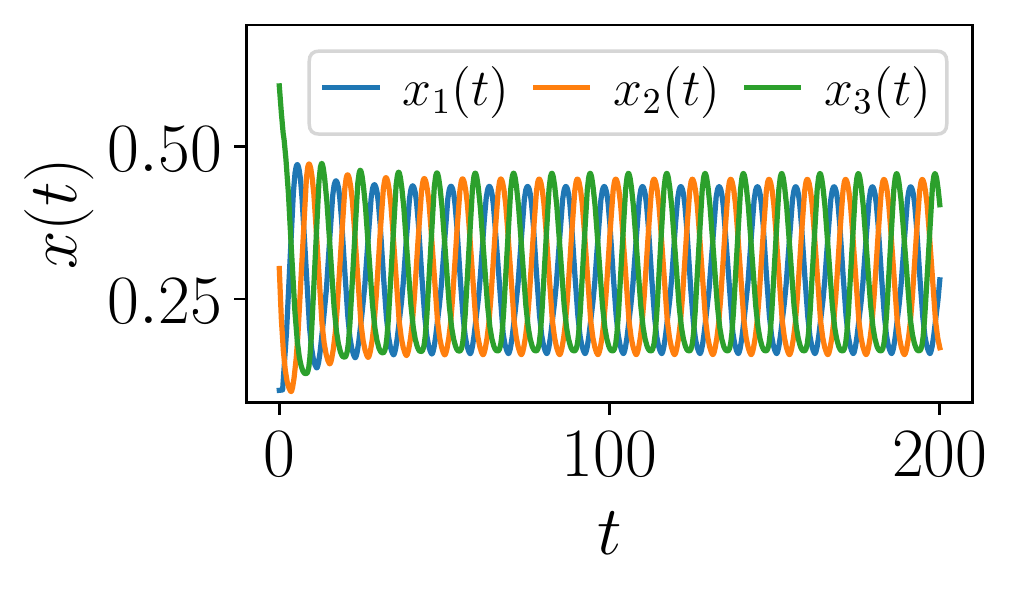}
  }
  \subfigure[Population in transition $y(t)$]{
    \includegraphics[trim={.1in .1in .1in .0in}, clip, width=1.55in]{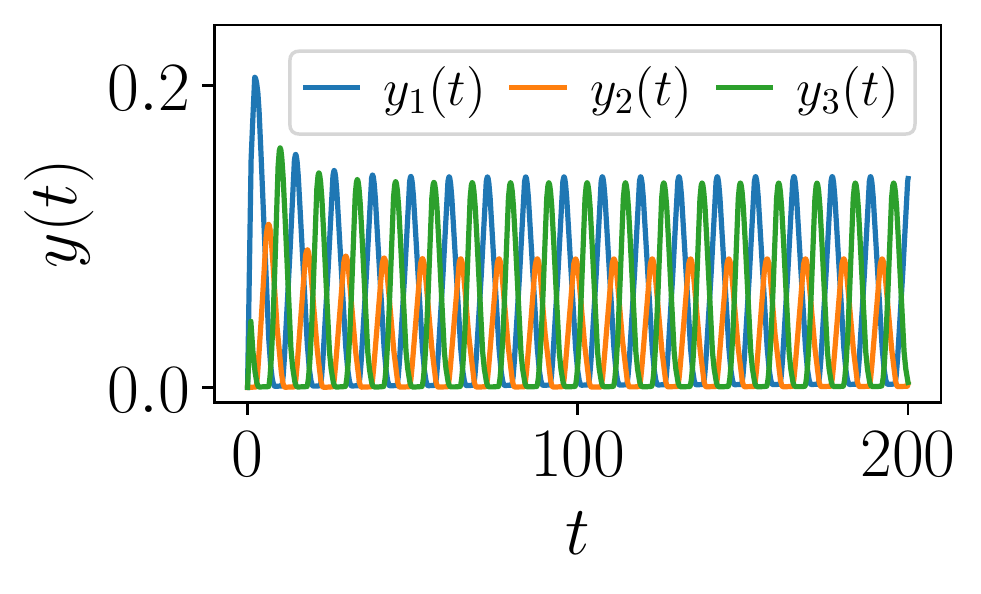}
  }
  
  \caption{Graphs depicting the trajectories of the (a) population state $x(t)$ and (b) population in transition $y(t)$ derived by EDM \eqref{eq:edm_transition_delay} ($d_{ij} = |i-j|, \varrho = 1/4, \lambda = 1$) in the RPS game \eqref{eq:payoff_function_rps} ($a=1, b=2$).}
  \label{fig:simulation_01}
\end{figure}
\begin{figure}
  \center
  \subfigure[Population state $x(t)$]{
    \includegraphics[trim={.1in .1in .1in 0in}, clip, width=1.55in]{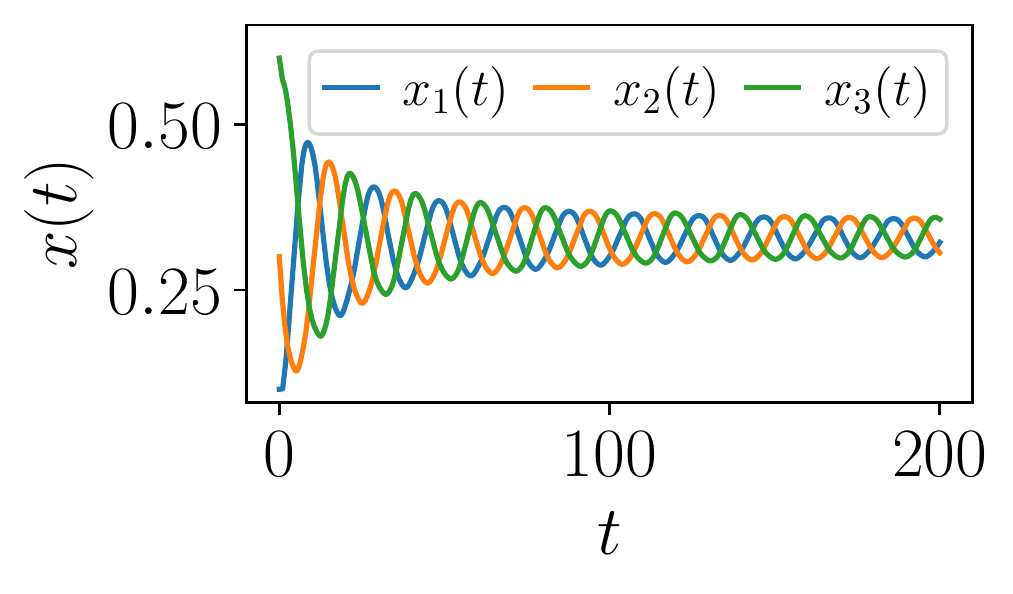}
  }
  \subfigure[Population in transition $y(t)$]{
    \includegraphics[trim={.1in .1in .1in .0in}, clip, width=1.55in]{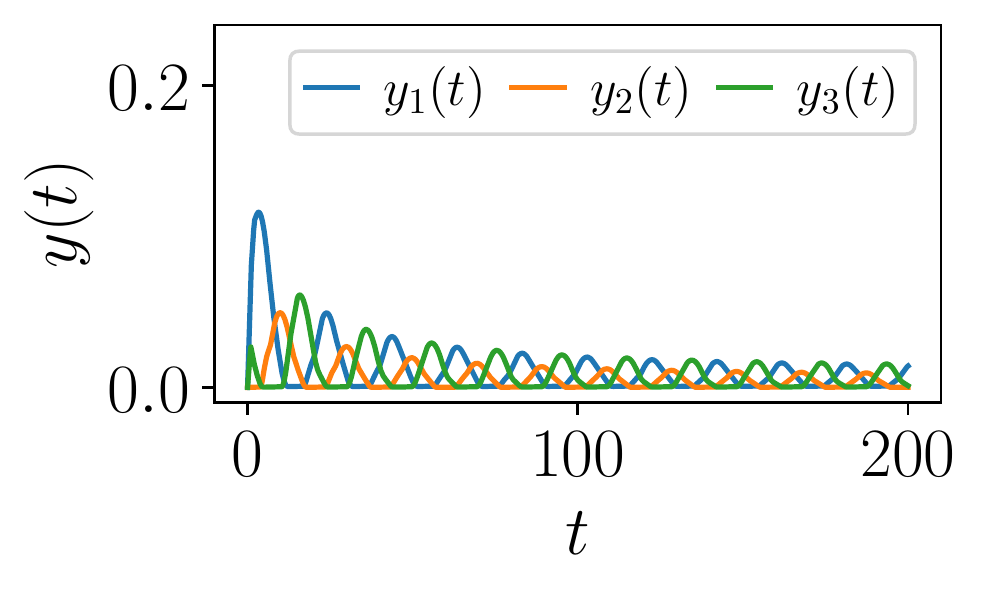}
  }

  \caption{Graphs depicting the trajectories of the (a) population state $x(t)$ and (b) population in transition $y(t)$ derived by EDM \eqref{eq:edm_transition_delay} ($d_{ij} = |i-j|, \varrho = 1/4, \lambda = 0.5$) in the RPS game \eqref{eq:payoff_function_rps} ($a=1, b=2$).}
  \label{fig:simulation_02}
\end{figure}

Let us define $y_i(t) = \sum_{j=1}^n y_{ji} (t)$ which denotes the portion of the population transitioning to strategy $i$. Then, $x_i(t) + y_i(t)$ quantifies the portion of the population revising their strategies to $i$ \textit{including} the agents in transition. Using \eqref{eq:original_edm}, we can describe the rate of change of $x_i(t) + y_i(t)$ by the following state equation.
\begin{align} \label{eq:edm_x_y}
  \dot x_{i} (t) + \dot y_i(t) = \mathcal V_i (x(t), p(t)),
\end{align}
where the vector field $\mathcal V_i$ is defined in \eqref{eq:original_edm}. Hence, from \eqref{eq:y_ij_dynamics} and \eqref{eq:edm_x_y}, we can derive the state equation for $x_i(t)$ as
\begin{multline} \label{eq:edm_transition_delay}
  \dot x_{i} (t) = \lambda \big( \textstyle\sum_{j=1}^n x_j(t \!-\! d_{ji}) \varrho \left[ p_i(t \!-\! d_{ji}) - p_j(t \!-\! d_{ji}) \right]_+ \\ - x_i(t) \textstyle\sum_{j=1}^n \varrho \left[ p_j(t) - p_i(t) \right]_+ \big), ~ t \geq d_{\text{max}},
\end{multline}
where $d_{\text{max}} = \max_{i,j \in \{1, \cdots, n\}} d_{ji}$.
Note that different from EDM \eqref{eq:original_edm} in its original form, when the agents' strategy revision is subject to time delays, the revised model \eqref{eq:edm_transition_delay} depends not only on the current information $x(t), p(t)$ but also on the past information ${x(t-d_{ji}), p(t-d_{ji})}$ about the population game.

Figs.~\ref{fig:simulation_01} and \ref{fig:simulation_02} depict the population state trajectories of \eqref{eq:edm_transition_delay} in the RPS game, explained in Example~\ref{example:rps_game}, with two different choices of the strategy revision rate ($\lambda = 0.5, 1$). We observe that in both cases, the  trajectories oscillate around the Nash equilibrium, but with smaller $\lambda$, both the level of the oscillation and the portion of the population in transition
decrease. Inspired by the observation, 
we investigate how the strategy revision rate can be tuned to achieve the convergence of the population state to the Nash equilibrium set.

We summarize the main problem of this paper as follows.
\noindent \rule{\columnwidth}{1pt}
\begin{problem} \label{problem:main}
  Design an algorithm to tune the strategy revision rate $\lambda$ of \eqref{eq:edm_transition_delay} and to achieve the convergence of the population state $x(t)$ to the Nash equilibrium set $\mathbb {NE} (\mathcal F)$ of an underlying contractive game $\mathcal F$:
  \begin{align} \label{eq:convergence_to_NE}
    \lim_{t \to \infty} \inf_{z \in \mathbb {NE}(\mathcal F)} \left\| x(t) - z \right\|_2 = 0.
  \end{align}
\end{problem}
\noindent \rule{\columnwidth}{1pt}

Note that according to Definition~\ref{definition:NE}, since every Nash equilibrium $z^{\footnotesize \text{NE}}$ in $\mathbb{NE} (\mathcal F)$ belongs to $\mathbb X$,
\eqref{eq:convergence_to_NE} requires the portion of the population in transition to vanish, i.e., 
\begin{align*}
  \textstyle\lim_{t \to \infty} \textstyle\sum_{i=1}^n y_i(t) = 1 - \lim_{t \to \infty} \textstyle\sum_{i=1}^n x_i(t) = 0.
\end{align*}
In this work, we focus on the case where the strategy revision rate $\lambda$ is homogeneous for all the transition between strategies. 
Our analysis is rooted in the notion of $\delta$-passivity for the Smith EDM \eqref{eq:original_edm}. As illustrated in \cite{10018262}, adopting heterogeneous strategy revision rates would not ensure the $\delta$-passivity of \eqref{eq:original_edm} and hence our proposed solution may not guarantee the convergence to the Nash equilibrium set. We leave as a future plan to further investigate such heterogeneous rate case.

\section{Tuning Strategy Revision Rate} \label{section:revision_rate_scheduling}
We describe an algorithm that iteratively updates $\lambda$ and in Section~\ref{section:analysis}, we explain how a sequence of the revision rates generated by the algorithm ensures the convergence of the population state to the Nash equilibrium set.

To explain the algorithm, we need the following lemma. The proof is given in Appendix of \cite{https://doi.org/10.48550/arxiv.2210.05472}.
\begin{lemma} \label{lemma:bounds_on_x_y}
  Given a fixed strategy revision rate $\lambda$, the states $x(t)$ and $y(t)$ satisfy
  \begin{align} \label{eq:bounds_on_x_y}
    \left\| \dot x(t) \right\|_2 \leq N \lambda, \quad 
    \left\| \dot y(t) \right\|_2 \leq M \lambda^2, ~ \forall t \geq d_{\text{max}},
  \end{align}
  where $d_{\text{max}} = \max_{i,j \in \{1, \cdots, n\}} d_{ji}$ and $N, M$ are positive constants.
\end{lemma}

The inequalities in \eqref{eq:bounds_on_x_y} imply that as $\lambda$ becomes smaller, i.e., the agents are less frequently revising their strategies, the rate of change of $y(t)$ decreases faster than that of $x(t)$. 
Therefore, the implication of Lemma~\ref{lemma:bounds_on_x_y} is that $\lambda$ can be used as a \textit{dial} to reduce the number of agents in transition while allowing them to have flexibility in switching between strategies. 

In what follows, we discuss tuning of $\lambda$.
Let us define $\bar y(t) = \lambda^{-2} y(t)$, $\bar{\mathcal V}(x(t), p(t)) = \lambda^{-1} \mathcal V (x(t), p(t))$, and $\bar{\mathcal S}(x(t), p(t)) = \lambda^{-1} \mathcal S (x(t), p(t))$. Using Lemma~\ref{lemma:bounds_on_x_y}, we derive the rate of change of $\bar{\mathcal S}$ along a solution to \eqref{eq:edm_transition_delay} as follows.
\begin{align} \label{eq:storage_function_upper_bound}
  &\frac{\mathrm d}{\mathrm d t} \bar{\mathcal S} \left( x(t), p(t) \right) \nonumber \\
  &= \lambda \Big( \bar{\mathcal V}^T (x(t), p(t)) D \mathcal F ( x(t) ) \bar{\mathcal V} (x(t), p(t)) \nonumber \\
  &\qquad\qquad + \nabla_x^T \bar{\mathcal S} (x(t), p(t)) \bar{\mathcal V}(x(t), p(t)) \nonumber \\ 
  &\qquad\qquad - \lambda \dot{\bar y}^T (t) \big( D \mathcal F^T (x(t)) \bar{\mathcal V} (x(t), p(t)) \nonumber \\
  &\qquad\qquad\qquad\qquad\qquad + \nabla_x \bar{\mathcal S} (x(t), p(t)) \big) \Big) \nonumber \\
  &\leq \lambda \Big(
    \nabla_x^T \bar{\mathcal S} (x(t), p(t)) \bar{\mathcal V}(x(t), p(t)) \nonumber \\ 
  &\qquad\qquad + \lambda M \big( B_{D \mathcal F} \left\| \bar{\mathcal V} (x(t), p(t)) \right\|_2 \nonumber \\
  &\qquad\qquad\qquad\qquad + \left\| \nabla_x \bar{\mathcal S} (x(t), p(t)) \right\|_2 \big) \Big),
\end{align}
where $p(t) = \mathcal F(x(t))$ and we use \eqref{eq:contrative_game_DF}, \eqref{eq:edm_x_y}, and the fact that $\nabla_p \mathcal S(x,p) = \mathcal V(x,p)$, derived in \cite[Theorem~III.3]{park2018cdc}. 
For notational convenience, we adopt
\begin{align} \label{eq:f_x_p}
  f(x,p) = M \left( B_{D \mathcal F} \left\| \bar{\mathcal V} (x, p) \right\|_2 + \left\| \nabla_x \bar{\mathcal S} (x, p) \right\|_2 \right).
\end{align}
Note that $f(x,p)$ does not depend on the revision rate $\lambda$.

Given an initial revision rate $\lambda = \lambda_0$ at $t = 0$, the following algorithm is used to tune $\lambda$ at each discrete time $t_k$ and to generate a sequence $\{\lambda_k\}_{k=1}^\infty$.

\noindent \rule{\columnwidth}{1pt}
\noindent \textbf{Algorithm~$1$: }
Suppose the strategy revision rate was updated to $\lambda_k$ at time $t_k$. At time instant $t \geq t_k + 2 d_{\text{max}}$, where $d_{\text{max}} = \max_{i,j \in \{1, \cdots, n\}} d_{ji}$, update the revision rate as
\begin{align} \label{eq:lambda_update_rule}
  \lambda = -\frac{1}{2} \frac{\nabla_x^T \bar{\mathcal S} (x(t), p(t)) \bar{\mathcal V}(x(t), p(t))}{f(x(t), p(t))}
\end{align}
if the following two conditions hold:

\begin{enumerate} 
\item $f(x(t), p(t)) > 0$ \label{condition_1}
\item $\nabla_x^T \bar{\mathcal S} (x(t), p(t)) \bar{\mathcal V}(x(t), p(t)) + \lambda_k \frac{f(x(t), p(t))}{1 - \delta} \geq 0$ \label{condition_2}
\end{enumerate}
where $\delta$ is a constant satisfying $0 < \delta < 1/2$. Then, set $t_{k+1} = t$ and $\lambda_{k+1} = \lambda$.

\noindent \rule{\columnwidth}{1pt}

Note that according to (\ref{condition_1}) and (\ref{condition_2}), the update rule \eqref{eq:lambda_update_rule} finds $\lambda$ that minimizes the upper bound of $\frac{\mathrm d}{\mathrm dt} \bar{\mathcal S}(x(t), p(t))$ given as in the last inequality of \eqref{eq:storage_function_upper_bound}.

\begin{remark} [Decreasing strategy revision rates] \label{remark:decreasing_sequence}
  Note that if we select $\lambda$ according to \eqref{eq:lambda_update_rule}, by (\ref{condition_2}), it satisfies
  \begin{align} \label{eq:decreasing_lambda}
    \lambda
    \!=\! - \frac{1}{2} \frac{\nabla_x^T \bar{\mathcal S} (x(t), p(t)) \bar{\mathcal V}(x(t), p(t))}{f(x(t), p(t))} \!\leq\! \frac{1}{2}\frac{\lambda_k}{1-\delta} \!<\! \lambda_k,
  \end{align}
  where the last inequality holds since $0 < \delta < 1/2$. Therefore, Algorithm~1 generates a decreasing sequence $\{\lambda_k\}_{k=1}^\infty$ of the revision rates satisfying $\lim_{k \to \infty} \lambda_k = 0$. \hfill\QED
\end{remark}

\begin{remark} [Comment on the constraint $t \geq t_k + 2 d_{\text{max}}$] \label{remark:constraint_on_time}
  Under Algorithm~1, since $\lambda$ switches over time, the parameter $\lambda$ 
  becomes time-dependent 
  and the state equation \eqref{eq:y_ij_dynamics} for $y_{ji}(t)$ can be rewritten as
\begin{multline} \label{eq:y_ij_dynamics_time_dependent}
  \dot y_{ji} (t) = \lambda(t) x_j(t) \varrho \left[ p_i(t) - p_j(t) \right]_+ \\
  - \lambda(t\!-\!d_{ji}) x_j(t\!-\!d_{ji}) \varrho \left[ p_i(t\!-\!d_{ji}) \!-\! p_j(t\!-\!d_{ji}) \right]_+
\end{multline}
for $t \geq d_{ji}$, where $\lambda(t)$ is the revision rate at time $t$ defined as $\lambda(t) = \lambda_k$ if $t \in [t_k, t_{k+1})$. With the new representation \eqref{eq:y_ij_dynamics_time_dependent}, the derivation of similar inequalities as in \eqref{eq:bounds_on_x_y}, which are key in establishing our main result, would be technically difficult. To circumvent such difficulty, we require the update time $t$ to satisfy $t \geq t_k + 2 d_{\text{max}}$ so that $\lambda(t)$ and $\lambda(t-d_{ji})$ in \eqref{eq:y_ij_dynamics_time_dependent} satisfy $\lambda(t)= \lambda(t-d_{ji})=\lambda_k$
for $t \in [t_k + 2 d_{\text{max}}, t_{k+1})$. \hfill\QED
\end{remark}

\begin{remark} [Existence of $t$ satisfying (\ref{condition_1}) and (\ref{condition_2})] \label{remark:existence_of_t}
  There may be no time instant $t \geq t_k + 2d_{\text{max}}$ satisfying (\ref{condition_1}) and (\ref{condition_2}) and no further update on $\lambda$,
  which we consider as the termination of the algorithm. In what follows, we show that the population state converges to the Nash equilibrium set even if 
  the algorithm terminates after a finite number of $\lambda$-updates.

  \noindent\paragraph*{\bf Case~I} Suppose $f(x(t), p(t)) = 0, ~ \forall t \geq t_k + 2 d_{\text{max}}$. Then, from \eqref{eq:derivative_S_negative} and \eqref{eq:storage_function_upper_bound}, it holds that
  \begin{align*}
    \frac{\mathrm d}{\mathrm d t} \bar{\mathcal S} \left( x(t), p(t) \right) \leq \lambda_k \nabla_x^T \bar{\mathcal S} (x(t), p(t)) \bar{\mathcal V}(x(t), p(t)) \leq 0.
  \end{align*}
  Noting that
  \begin{multline*}
    \nabla_x^T \bar{\mathcal S} (x(t), p(t)) \bar{\mathcal V}(x(t), p(t)) = 0 \\\iff \bar{\mathcal S}(x(t), p(t)) = 0,
  \end{multline*}
  we conclude $\lim_{t \to \infty} \bar{\mathcal S}(x(t), p(t)) = 0$.

  \noindent\paragraph*{\bf Case~II} Suppose the following inequality holds:
  \begin{multline}
    \nabla_x^T \bar{\mathcal S} (x(t), p(t)) \bar{\mathcal V}(x(t), p(t)) + \lambda_k \frac{f(x(t), p(t))}{1-\delta} < 0, \\ ~ \forall t \geq t_k + 2 d_{\text{max}}.
  \end{multline}
  Then, from \eqref{eq:derivative_S_negative} and \eqref{eq:storage_function_upper_bound}, it holds that
  \begin{align*}
    \frac{\mathrm d}{\mathrm d t} \bar{\mathcal S} \left( x(t), p(t) \right) \leq \delta \lambda_k \nabla_x^T \bar{\mathcal S} (x(t), p(t)) \bar{\mathcal V}(x(t), p(t)) \leq 0.
  \end{align*}
  By the same argument as in (\textbf{Case~I}), we have $\lim_{t \to \infty} \bar{\mathcal S}(x(t), p(t)) = 0$.

  For both (\textbf{Case~I}) and (\textbf{Case~II}), by \eqref{eq:S_convergence}, it holds that $\lim_{t \to \infty} x_{i}(t) \varrho [p_j(t) - p_i(t)]_+ = 0,~ \forall i,j \in \{1, \cdots, n\}$. Then, using Lemma~\ref{lemma:convergence_revision_protocol} stated in Section~\ref{section:analysis}, 
  we conclude $\lim_{t \to \infty} \inf_{z \in \mathbb{NE}(\mathcal F)} \|x(t) - z\|_2 = 0$. \hfill\QED

\end{remark}

\subsection{Convergence Analysis} \label{section:analysis}
Recall that by Remark~\ref{remark:existence_of_t}, even if Algorithm~1 terminates after a finite number of $\lambda$-updates, the population state still converges to the Nash equilibrium set. In this section, we consider the case where Algorithm~1 generates an infinite sequence $\{\lambda_k\}_{k=1}^\infty$ of strategy revision rates.
The following theorem states the main result. 
\begin{theorem} \label{theorem:convergence_result}
  Let $\{\lambda_k\}_{k=1}^\infty$ be a decreasing sequence  of strategy revision rates determined by Algorithm~1. The population state converges to the Nash equilibrium set, i.e., it holds that
  \begin{align} 
    \lim_{t \to \infty} \inf_{z \in \mathbb{NE}(\mathcal F)} \left\| x(t) - z \right\|_2 = 0.
  \end{align}
\end{theorem}

Using the following two lemmas, we prove the theorem by showing that the $\delta$-storage function $\mathcal S$ decreases at each time instant $t_k$ of the $\lambda$-update (Lemma~\ref{lemma:bound_on_S}) which implies that the population state converges to the Nash equilibrium set $\mathbb{NE}(\mathcal F)$ (Lemma~\ref{lemma:convergence_revision_protocol}).
We provide the proofs of the theorem and lemmas in Appendix of \cite{https://doi.org/10.48550/arxiv.2210.05472}.
\begin{lemma} \label{lemma:bound_on_S}
  Given $\mathcal V = (\mathcal V_1, \cdots, \mathcal V_n)$ and $\mathcal S$ as in \eqref{eq:original_edm} and \eqref{eq:smith_storage_function}, respectively, define
  $\bar{\mathcal V} = \lambda^{-1} \mathcal V$ and $\bar{\mathcal S} = \lambda^{-1} \mathcal S$. Let $K$ be a positive constant satisfying $K > (1 - \delta)^{-1} f(x, p)$ for all $x$ in $\mathbb X^{\footnotesize \text{ext}}$ and $p = \mathcal F(x)$, where $\delta$ and $f(x,p)$ are defined in Algorithm~1.
  There is a decreasing function $\epsilon: \mathbb R_+ \to \mathbb R_+$ satisfying
  \begin{align} \label{eq:bound_on_S}
    \nabla_x^T \bar{\mathcal S} (x, p) \bar{\mathcal V}(x, p) + \lambda K > 0 \implies \bar{\mathcal S}(x,p) < \epsilon (\lambda)
  \end{align}
  and 
  $\lim_{\lambda \to 0} \epsilon (\lambda) = 0$.
\end{lemma}

\begin{lemma} \label{lemma:convergence_revision_protocol}
  The following statement is true.
  \begin{multline}
    \lim_{t \to \infty} x_{i}(t) \varrho [p_j(t) - p_i(t)]_+ = 0,~ \forall i,j \in \{1, \cdots, n\} \\ \implies \lim_{t \to \infty} \inf_{z \in \mathbb{NE} (\mathcal F)} \left\| x(t) - z \right\|_2 = 0.
  \end{multline}
\end{lemma}

\section{Simulations} \label{section:simulations}
\begin{figure}
  \center
  \subfigure[Population state $x(t)$]{
    \includegraphics[trim={.1in .1in .1in .0in}, clip, width=1.55in]{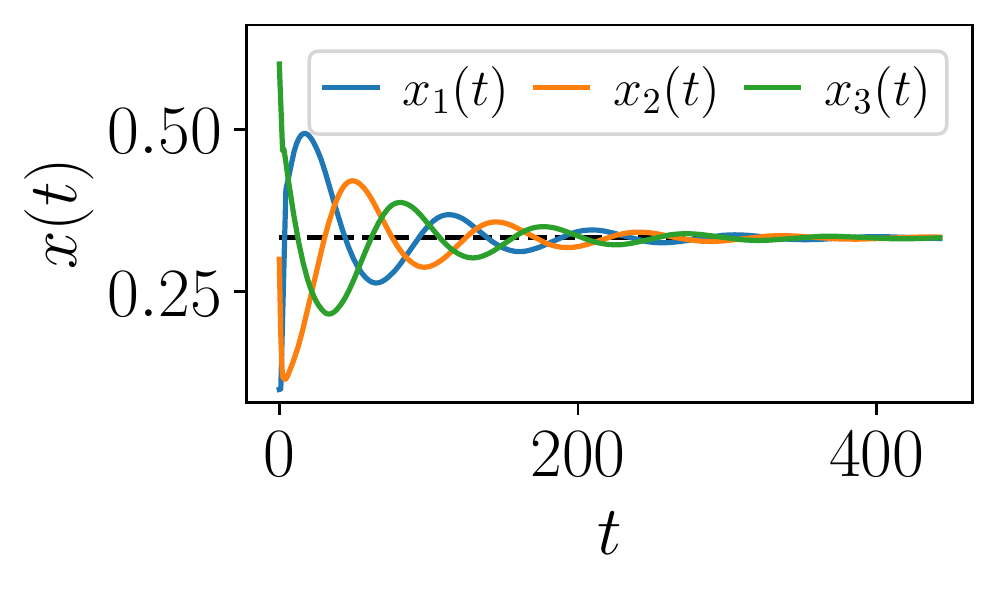}
  }
  \subfigure[Population in transition $y(t)$]{
    \includegraphics[trim={.1in .1in .1in .0in}, clip, width=1.55in]{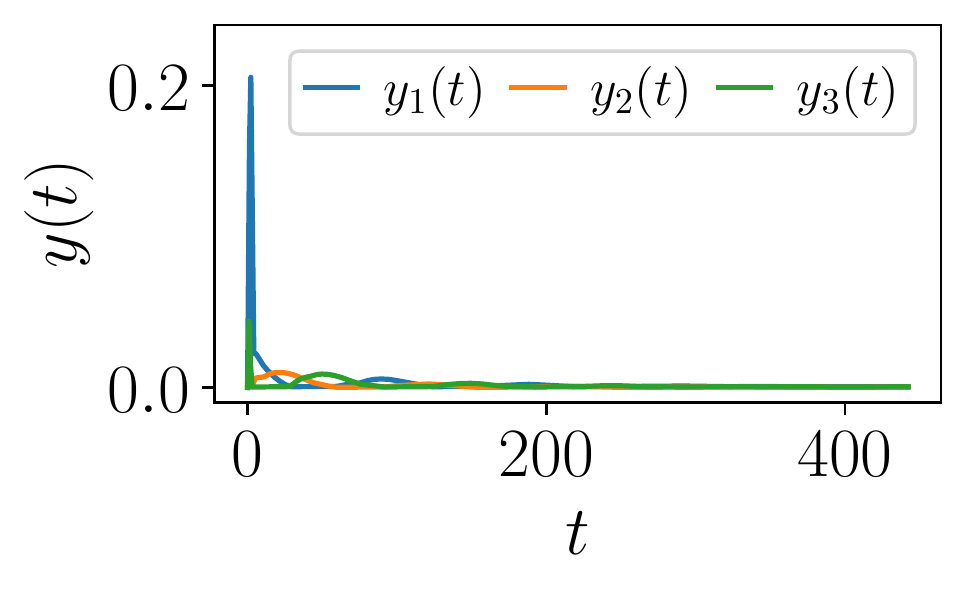}
  }

  \caption{Graphs illustrating the trajectories of the (a) population state $x(t)$ and (b) population in transition $y(t)$ derived by EDM \eqref{eq:edm_transition_delay} ($d_{ij} = |i-j|, \varrho = 1/4$) in the RPS game \eqref{eq:payoff_function_rps} ($a=1, b=2$).}
  \label{fig:simulation_03}
  \vspace{-4ex}
\end{figure}

Using Example~\ref{example:rps_game}, we illustrate our convergence result (Theorem~\ref{theorem:convergence_result}) through simulations. To this end, we adopt the payoff function \eqref{eq:payoff_function_rps} with $a=1, b=2$ as follows.
\begin{align} \label{eq:payoff_function_rps_simulation}
  \mathcal F(x) =
  \begin{pmatrix}
    - x_2 + 2 x_3 \\
    2 x_1 -  x_3 \\
    - x_1 + 2 x_2
  \end{pmatrix}
\end{align}
Note that $\mathcal F$ is contractive satisfying \eqref{eq:contrative_game_DF} and the upper bound of $D \mathcal F$ can be computed as $B_{D\mathcal F} \approx 2.65$.

We use Algorithm~1 to compute the sequence $\{ \lambda_k \}_{k=1}^\infty$ for EDM~\eqref{eq:edm_transition_delay} with the initial revision rate $\lambda_0=1$. Fig.~\ref{fig:simulation_03} depicts the trajectories for both $x(t)$ and $y(t)$. Contrast to the fixed revision rate cases illustrated in Figs.~\ref{fig:simulation_01} and \ref{fig:simulation_02}, through tuning the revision rate using Algorithm~1, we observe that the population state $x(t)$ converges to the Nash equilibrium $(1/3, 1/3, 1/3)$ of \eqref{eq:payoff_function_rps_simulation}. Also the portion $y(t)$ of the population in transition vanishes as the revision rate decreases.

\section{Conclusions} \label{section:conclusions}
We investigate a multi-agent decision problem in population games where the strategy revision of the agents is subject to time delays, which prevent the state of the population from converging to the Nash equilibrium set. We propose an algorithm that judiciously decreases the rate of the agents' strategy revision and prove that by tuning the revision rate, the population state converges to the Nash equilibrium set when the underlying population game is contractive.

As future directions, we plan to improve the proposed algorithm to attain faster convergence to the Nash equilibrium set. The current version of the algorithm is designed to minimize the upper bound of the rate of change of the $\delta$-passive storage function; however this is mainly used to establish the convergence result and would not guarantee fast convergence.
Also, validating the proposed approach in engineering applications, where the population would consist of a finite number of decision-making agents, will be another direction we plan to explore.

\appendix
\subsection{Proof of Lemma~\ref{lemma:bounds_on_x_y}}
Using
\eqref{eq:edm_transition_delay}, we can derive a bound on $\dot x_i(t)$ as follows. 
\begin{align} \label{eq:bound_on_x_i_dot}
  \left| \dot x_i(t) \right|
  &\leq \lambda \Big( \textstyle\sum_{j=1}^n x_{j}(t-d_{ji}) \varrho \left[ p_i(t-d_{ji}) - p_j(t-d_{ji}) \right]_+ \nonumber \\
  &\qquad\qquad + x_i(t) \textstyle\sum_{j=1}^n  \varrho \left[ p_j(t) - p_i(t) \right]_+ \Big) \nonumber \\
  &\leq \lambda ( n-1 + x_i(t) ) \leq n \lambda,
\end{align}
where to establish the second inequality, we use \eqref{eq:revision_protocol_inequality} and the fact that 
\begin{subequations}
  \begin{align*}
    x_{j}(t-d_{ji}) \varrho \left[ p_i(t-d_{ji}) - p_j(t-d_{ji}) \right]_+ &\leq 1, ~ j \neq i, \\
    x_{j}(t-d_{ji}) \varrho \left[ p_i(t-d_{ji}) - p_j(t-d_{ji}) \right]_+ &= 0, ~ j = i.
  \end{align*}
\end{subequations}
Using \eqref{eq:bound_on_x_i_dot}, 
we can establish Lipschitz continuity of $x(t)$: For any $d \geq 0$, 
it holds that
\begin{align} \label{eq:x_j_Lipschitz_continuity}
  \left| x_j(t) - x_j(t-d) \right| \leq \max_{\tau \in [t-d, t]} \left| \dot x_j (\tau) \right| d \leq n d \lambda
\end{align}
and
\begin{align} \label{eq:x_Lipschitz_continuity}
  \left\| x(t) - x(t-d) \right\|_2 
  \leq n^{1.5} d \lambda.
\end{align}
Consequently, using \eqref{eq:x_Lipschitz_continuity}, the revision protocol $\varrho [ p_i(t) - p_j(t) ]_+$ satisfies
\begin{align} \label{eq:revision_protocol_Lipschitz_continuity}
  &\left| \varrho \left[ p_i(t) - p_j(t) \right]_+ - \varrho \left[ p_i(t-d_{ji}) - p_j(t-d_{ji}) \right]_+ \right| \nonumber \\
  &\leq \varrho \left| p_i(t) - p_j(t)  -  p_i(t-d_{ji}) + p_j(t-d_{ji}) \right| \nonumber \\
  &\leq \varrho \left| p_i(t) -  p_i(t-d_{ji}) \right| + \varrho \left| p_j(t) - p_j(t-d_{ji}) \right| \nonumber \\
  &\leq \varrho \left( \left\| D \mathcal F_i \right\|_2 + \left\| D \mathcal F_j \right\|_2 \right) \left\| x(t) - x(t-d_{ji}) \right\|_2 \nonumber \\
  &\leq 2 \varrho B_{D \mathcal F} n^{1.5} d_{ji} \lambda,
\end{align}
where $\mathcal F_i$ is the $i$-th component of $\mathcal F = (\mathcal F_1, \cdots, \mathcal F_n)$, $\left\| D \mathcal F_i \right\|_2 = \max_{x \in \mathbb X^{\text{ext}}} \left\| D \mathcal F_i (x) \right\|_2 \leq B_{D \mathcal F}$, and we use
\begin{align}
  \left| p_i(t) -  p_i(t-d_{ji}) \right|
  &= \left| \mathcal F_i(x(t)) -  \mathcal F_i(x(t-d_{ji})) \right| \nonumber \\
  &\leq \|D \mathcal F_i\|_2 \left\| x(t) - x(t-d_{ji}) \right\|_2 \nonumber \\
  &\leq B_{D \mathcal F} n^{1.5} d_{ji} \lambda.
\end{align}
Therefore, using \eqref{eq:x_j_Lipschitz_continuity} and \eqref{eq:revision_protocol_Lipschitz_continuity}, we derive a bound on $\dot y(t)$ as follows.
\begin{align} \label{eq:bound_on_y_i_dot}
  &\left| \dot y_i(t) \right| \nonumber \\
  &\leq \lambda \sum_{j=1}^n \Big| x_j(t) \varrho \left[ p_i(t) - p_j(t) \right]_+ \nonumber \\
  &\qquad\qquad - x_j(t-d_{ji}) \varrho \left[ p_i(t-d_{ji}) - p_j(t-d_{ji}) \right]_+ \Big| \nonumber \\
  &\leq \lambda \sum_{j=1}^n \Big| x_j(t) \varrho \left[ p_i(t) - p_j(t) \right]_+ \nonumber \\
  &\qquad\qquad - x_j(t-d_{ji}) \varrho \left[ p_i(t) - p_j(t) \right]_+ \Big| \nonumber \\
  &\quad + \lambda \sum_{j=1}^n \Big| x_j(t-d_{ji}) \varrho \left[ p_i(t) - p_j(t) \right]_+ \nonumber \\
  &\quad\qquad\qquad - x_j(t-d_{ij}) \varrho \left[ p_i(t-d_{ji}) - p_j(t-d_{ji}) \right]_+ \Big| \nonumber \\
  &\leq \lambda \sum_{j=1}^n n d_{ji} \lambda \varrho \left[ p_i(t) - p_j(t) \right]_+ + \lambda \sum_{j=1}^n 2 \varrho B_{D \mathcal F} n^{1.5} d_{ji} \lambda \nonumber \\
  &\leq (1 + 2 \varrho B_{D \mathcal F} n^{0.5}) n d_{i} \lambda^2
\end{align}
where $d_i = \sum_{j=1}^n d_{ji}$. Hence, from \eqref{eq:bound_on_x_i_dot} and \eqref{eq:bound_on_y_i_dot}, we conclude that
\begin{align*}
  \left\| \dot x(t) \right\|_2 \leq N\lambda, \quad \left\| \dot y(t) \right\|_2 \leq M \lambda^2
\end{align*}
where $N = n^{1.5}$ and $M = (1 + 2 \varrho B_{D \mathcal F} n^{0.5}) n \sqrt{\sum_{i=1}^n d_i^2}$. This completes the proof.
\hfill\QED

\subsection{Proof of Lemma~\ref{lemma:bound_on_S}}
We first show that there is $K$ such that $K > (1-\delta)^{-1} f(x,p)$ holds for all $x$ in $\mathbb X^{\footnotesize \text{ext}}$ and $p = \mathcal F(x)$, where $f(x,p)$ is defined in \eqref{eq:f_x_p}.
Note that using \eqref{eq:revision_protocol_inequality}, we have
\begin{multline} \label{eq:bound_on_V_i_bar}
  \left| \bar{\mathcal V}_i (x,p) \right| = \Big| \textstyle\sum_{j=1}^n x_j \varrho \left[ p_i - p_j \right]_+ \\
  - x_i \textstyle\sum_{j=1}^n \varrho \left[ p_j - p_i \right]_+ \Big| \leq 1
\end{multline}
and
\begin{align} \label{eq:partial_derivative_S_x}
  \left| \frac{\partial \bar{\mathcal S}}{\partial x_k} (x, p) \right|
  = \frac{1}{2} \sum_{j=1}^n \varrho [p_j - p_k]_+^2 \leq \frac{1}{2 \varrho}.
\end{align}

Therefore, using the definition of $f(x,p)$ and \eqref{eq:bound_on_V_i_bar}, \eqref{eq:partial_derivative_S_x}, we can conclude
\begin{align}
  &(1-\delta)^{-1} f(x,p) \nonumber \\
  &= (1-\delta)^{-1} M \left( B_{D \mathcal F} \left\| \bar{\mathcal V} (x, p) \right\|_2 + \left\| \nabla_x \bar{\mathcal S} (x, p) \right\|_2 \right) \nonumber \\
  &\leq (1-\delta)^{-1} M \left( B_{D\mathcal F} + \frac{1}{2 \varrho} \right) n^{0.5} < K
\end{align}
if we choose $K > (1-\delta)^{-1} M \left( B_{D\mathcal F} + \frac{1}{2 \varrho} \right) n^{0.5}$.

To conclude the proof, we prove the contrapositive of \eqref{eq:bound_on_S}:
\begin{align}
  \bar{\mathcal S}(x,p) \geq \epsilon (\lambda) \implies
  \nabla_x^T \bar{\mathcal S} (x, p) \bar{\mathcal V}(x, p) + K \lambda \leq 0
\end{align}
Suppose $\bar{\mathcal S}(x, p) \geq \epsilon$ then according to \eqref{eq:smith_storage_function}, it holds that
\begin{align}
  \max_{1 \leq i,j \leq n} \frac{\varrho}{2} x_i [p_j - p_i]_+^2 \geq \frac{\epsilon}{n^2}.
\end{align}
Otherwise $\bar{\mathcal S}(x,p) < \epsilon$ which is a contradiction.
Let $i^\ast, j^\ast$ be indices for which the following holds.
\begin{align}
  \max_{1 \leq i,j \leq n} \frac{\varrho}{2} x_i [p_j - p_i]_+^2
  &= \frac{\varrho}{2} x_{i^\ast} [p_{j^\ast} - p_{i^\ast}]_+^2
\end{align}
from which we obtain
\begin{align}
  \frac{\epsilon}{n^2} \leq \max_{1 \leq i,j \leq n} \frac{\varrho}{2} x_i [p_j - p_i]_+^2
  &= \frac{\varrho}{2} x_{i^\ast} [p_{j^\ast} - p_{i^\ast}]_+^2 \nonumber \\
  &\leq \frac{1}{2} [p_{j^\ast} - p_{i^\ast}]_+,
\end{align}
where we use \eqref{eq:revision_protocol_inequality} to establish the last inequality. Therefore, we have that
\begin{subequations} \label{eq:lower_bound_on_revision_protocol}
  \begin{align}
    x_{i^\ast} [p_{j^\ast} - p_{i^\ast}]_+^2 &\geq \frac{2 \epsilon}{\varrho n^2} \\
    [p_{j^\ast} - p_{i^\ast}]_+ &\geq \frac{2 \epsilon}{n^2}.
  \end{align}
\end{subequations}

Also we can derive
\begin{align}
  &\nabla_x^T \bar{\mathcal S} (x, p) \bar{\mathcal V}(x, p) \nonumber \\
  &= \sum_{i=1}^n \sum_{j=1}^n \big( x_j \varrho \left[ p_i \!-\! p_j \right]_+ \!-\! x_i \varrho \left[ p_j \!-\! p_i \right]_+ \big)  \frac{\varrho}{2} \sum_{k=1}^n [p_k \!-\! p_i]_+^2 \nonumber \\
  &= \frac{\varrho^2}{2} \sum_{i=1}^n \sum_{j=1}^n \Big( x_j \left[ p_i \!-\! p_j \right]_+ \sum_{k=1}^n \left( [p_k \!-\! p_i]_+^2 \!-\! [p_k \!-\! p_j]_+^2\right) \Big) \nonumber \\
  &\leq \frac{\varrho^2}{2} \Big( x_j \left[ p_i - p_j \right]_+ \left( [p_k - p_i]_+^2 - [p_k - p_j]_+^2\right) \Big)
\end{align}
for any $i,j,k \in \{1, \cdots, n\}$. If we choose $i = k = j^\ast, j = i^\ast$, using \eqref{eq:lower_bound_on_revision_protocol}, we can derive
\begin{align}
  \nabla_x^T \bar{\mathcal S} (x, p) \bar{\mathcal V}(x, p)
  &\leq -\frac{\varrho^2}{2} x_{i^\ast} \left[ p_{j^\ast} - p_{i^\ast} \right]_+ \left[ p_{j^\ast} - p_{i^\ast} \right]_+^2 \nonumber \\
  &\leq - \frac{2 \epsilon^2 \varrho}{n^4}
\end{align}

Let us assign $\epsilon(\lambda) = \sqrt{\frac{n^4 K \lambda}{2 \varrho}}$, then we can conclude
\begin{align*}
  \nabla_x^T \bar{\mathcal S} (x, p) \bar{\mathcal V}(x, p) + K \lambda \leq 0
\end{align*}
and also we can verify that $\epsilon(\lambda) \to 0$ as $\lambda \to 0$. \hfill \QED

\subsection{Proof of Lemma~\ref{lemma:convergence_revision_protocol}}

Suppose $y_{ji}(0) = 0, ~ \forall i,j \in \{1, \cdots, n\}$, i.e., there is no agent in transition for the strategy revision at the beginning of the game. Let $t_2$ be the time instant that Algorithm~1 updates the revision rate for the second time. For $t \geq t_2$, a solution $y_{ji}(t)$ to \eqref{eq:y_ij_dynamics_time_dependent} 
satisfies
\begin{align} \label{eq:bound_on_y_ji}
  &y_{ji}(t) \nonumber \\
  &= \int_{0}^t \big( \lambda(\tau) x_j(\tau) \varrho \left[ p_i(\tau) - p_j(\tau) \right]_+ \nonumber \\
  &\qquad \!-\! \lambda(\tau \!-\! d_{ji}) x_j (\tau \!-\! d_{ji}) \varrho \left[ p_i (\tau \!-\! d_{ji}) \!-\! p_j(\tau \!-\! d_{ji}) \right]_+ \big) \mathrm d\tau \nonumber \\
  &= \int_{t-d_{ji}}^t \lambda(\tau) x_j(\tau) \varrho \left[ p_i(\tau) - p_j(\tau) \right]_+ \, \mathrm d\tau \nonumber \\
  &\qquad \!-\! \int_{-d_{ji}}^{0} \lambda(\tau) x_j(\tau) \varrho \left[ p_i(\tau) - p_j(\tau) \right]_+ \, \mathrm d\tau \nonumber \\
  &\leq \lambda_1 \int_{t-d_{ji}}^t x_j(\tau) \varrho \left[ p_i(\tau) - p_j(\tau) \right]_+ \, \mathrm d\tau,
\end{align}
where we use 
the facts that
\begin{align*}
  \int_{-d_{ji}}^{0} \lambda(\tau) x_j(\tau) \varrho \left[ p_i(\tau) - p_j(\tau) \right]_+ \, \mathrm d\tau \geq 0,
\end{align*}
$\lambda(\tau) = \lambda_k \text{ if } \tau \in [t_k, t_{k+1})$ with $t_{k+1} \geq t_{k} + 2 d_{\text{max}}$, and $\{\lambda_k\}_{k=1}^\infty$ is a decreasing sequence.

Suppose
\begin{align} \label{eq:vanishing_revision_protocol}
  \lim_{t \to \infty} x_{i}(t) \varrho [p_j(t) - p_i(t)]_+ = 0,~ \forall i,j \in \{1, \cdots, n\}
\end{align}
holds as in the statement of the lemma. Then, from \eqref{eq:bound_on_y_ji}, we have that
\begin{align} \label{eq:vanishing_y_ji}
  \lim_{t \to \infty} y_{ji}(t) = 0, ~ \forall i,j \in \{1, \cdots, n\}.
\end{align}

To complete the proof, using \eqref{eq:vanishing_revision_protocol} and \eqref{eq:vanishing_y_ji}, we establish $\lim_{t \to \infty} \inf_{z \in \mathbb{NE}(\mathcal F)} \|x(t) - z\|_2 = 0$. By contradiction, suppose that there is a converging sequence $\{x(t_k)\}_{k=1}^\infty$ for which its limit point $x^\ast$ is not contained in $\mathbb{NE} (\mathcal F)$. By \eqref{eq:vanishing_revision_protocol} and \eqref{eq:vanishing_y_ji}, it holds that $\sum_{i=1}^n x_i^\ast = 1$ and
\begin{align*}
  x_i^\ast > 0 \implies \mathcal F_i(x^\ast) = \max_{1 \leq j \leq n} \mathcal F_j(x^\ast), ~\forall i \in \{1, \cdots, n\}.
\end{align*}
Hence, for any $z$ in $\mathbb X$, we have that
\begin{align*}
  \left( x^\ast - z \right)^T \mathcal F(x^\ast) = \max_{1 \leq j \leq n} \mathcal F_j (x^\ast) - z^T \mathcal F(x^\ast) \geq 0.
\end{align*}
According to Definition~\ref{definition:NE}, we conclude that $x^\ast$ is the Nash equilibrium, which is a contradiction. This completes the proof. \hfill \QED

\subsection{Proof of Theorem~\ref{theorem:convergence_result}}
Note that by (\ref{condition_2}) of Algorithm~1, at each 
update time $t_{k+1}$, it holds that
\begin{align*}
  &\nabla_x^T \bar{\mathcal S} (x(t_{k+1}), p(t_{k+1})) \bar{\mathcal V}(x(t_{k+1}), p(t_{k+1})) + \lambda_k K \\
  &> \nabla_x^T \bar{\mathcal S} (x(t_{k+1}), p(t_{k+1})) \bar{\mathcal V}(x(t_{k+1}), p(t_{k+1})) \\
  &\qquad + \lambda_k f(x(t_{k+1}),p(t_{k+1})) \geq 0,
\end{align*}
where $K$ is a constant satisfying $K > (1-\delta)^{-1} f(x,p) \geq f(x,p)$ as in the statement of Lemma~\ref{lemma:bound_on_S}. Invoking Lemma~\ref{lemma:bound_on_S}, there is a decreasing function $\epsilon: \mathbb R_+ \to \mathbb R_+$ satisfying $\bar{\mathcal S} (x(t_{k+1}),p(t_{k+1})) < \epsilon (\lambda_k)$. In what follows, we show that there is a function $\bar{\epsilon}: \mathbb R_+ \to \mathbb R_+$ satisfying
\begin{align}
  \bar{\mathcal S} (x(t), p(t)) < \bar{\epsilon} (\lambda_k), ~ \forall t \in [t_{k+1}, t_{k+2})
\end{align}
for every positive integer $k$ and $\lim_{k \to \infty} \bar{\epsilon} (\lambda_k) = 0$.

For $t \in [t_{k+1}, t_{k+1} + 2 d_{\text{max}})$, it holds that
\begin{align} \label{eq:bound_on_S_bar}
  &\bar{\mathcal S}(x(t), p(t)) - \bar{\mathcal S}(x(t_{k+1}), p(t_{k+1})) \nonumber \\
  &\leq \max_{x \in \mathbb X^{\footnotesize \text{ext}}} \left\| \nabla_x \bar{\mathcal S}(x, p) |_{p = \mathcal F(x)} \right\|_2 \left\| x(t) - x(t_{k+1}) \right\|_2  \nonumber \\
  &\qquad + B_{D\mathcal F}\max_{x \in \mathbb X^{\footnotesize \text{ext}}} \left\| \nabla_p \bar{\mathcal S}(x, p)|_{p = \mathcal F(x)} \right\|_2 \left\| x(t) - x(t_{k+1}) \right\|_2  \nonumber \\
  &\leq 2d_{\text{max}} L \max_{\tau \in [t_{k+1}, t_{k+1}+2d_{\text{max}})} \left\| \dot x(\tau) \right\|_2,
\end{align}
where $L$ is a constant satisfying\footnote{Note that such constant $L$ exists since $\mathcal S$ is continuously differentiable, $\mathcal F$ is continuous, and $\mathbb X^{\footnotesize \text{ext}}$ is compact.}
\begin{multline*}
  L = \max_{x \in \mathbb X^{\footnotesize \text{ext}}} \left\| \nabla_x \bar{\mathcal S}(x, p) |_{p = \mathcal F(x)} \right\|_2 \\ + B_{D\mathcal F}\max_{x \in \mathbb X^{\footnotesize \text{ext}}} \left\| \nabla_p \bar{\mathcal S}(x, p)|_{p = \mathcal F(x)} \right\|_2.
\end{multline*}

Note that when the revision rate is repeatedly updated by Algorithm~1, $\lambda$ becomes a time-dependent parameter and the state equation \eqref{eq:y_ij_dynamics} for $y_{ji}(t)$ can be rewritten as in  \eqref{eq:y_ij_dynamics_time_dependent}.
Hence, using \eqref{eq:edm_x_y} and \eqref{eq:y_ij_dynamics_time_dependent}, it holds that for $t \in [t_{k+1}, t_{k+2})$,
\begin{align} \label{eq:bound_on_x_i_dot_time_dependent_rate}
  &\left| \dot x_{i} (t) \right| \nonumber \\
  &\leq \textstyle\sum_{j=1}^n \lambda(t \!-\! d_{ji}) x_j(t \!-\! d_{ji}) \varrho \left[ p_i(t \!-\! d_{ji}) \!-\! p_j(t \!-\! d_{ji}) \right]_+ \nonumber \\
  & \qquad + \textstyle\sum_{j=1}^n \lambda(t) x_i(t) \varrho \left[ p_j(t) - p_i(t) \right]_+ \nonumber \\
  &\leq \textstyle\sum_{\substack{j=1 \\ j \neq i}}^n \lambda(t - d_{ji}) + \lambda_{k} \nonumber \\
  &\leq n \lambda_{k},
\end{align}
where we use \eqref{eq:revision_protocol_inequality}, and the facts that $\{ \lambda_k \}_{k=1}^\infty$ is a decreasing sequence and $t_{k+1} \geq t_k + 2 d_{\text{max}}$. Consequently, from \eqref{eq:bound_on_S_bar} and \eqref{eq:bound_on_x_i_dot_time_dependent_rate}, we can derive
\begin{multline}
  \bar{\mathcal S}(x(t), p(t)) 
  \leq \bar{\mathcal S}(x(t_{k+1}), p(t_{k+1})) + 2d_{\text{max}} L n^{1.5} \lambda_k, \\ ~\forall t \in [t_{k+1}, t_{k+1}+2d_{\text{max}}).
\end{multline}

On the other hand, for $t \in [t_{k+1} + 2 d_{\text{max}}, t_{k+2})$, at least one of  (\ref{condition_1}) and (\ref{condition_2}) of Algorithm~1 is violated and using \eqref{eq:derivative_S_negative} and \eqref{eq:storage_function_upper_bound}, we can derive
\begin{multline}
  \frac{\mathrm d}{\mathrm dt} \bar{\mathcal S}(x(t), p(t)) \leq \lambda_{k+1} \big( \nabla_x^T \bar{\mathcal S} (x(t), p(t)) \bar{\mathcal V}(x(t), p(t)) \\ + \lambda_{k+1} f(x(t), p(t)) \big) \leq 0.
\end{multline}
Hence, the function $\bar{\mathcal S}(x(t), p(t))$ does not increase over $[t_{k+1} + 2 d_{\text{max}}, t_{k+2})$.
Therefore, we conclude that for all $t \in [t_{k+1}, t_{k+2})$,
\begin{align}
  \bar{\mathcal S}(x(t), p(t)) 
  &\leq \bar{\mathcal S}(x(t_{k+1}), p(t_{k+1})) + 2d_{\text{max}} L n^{1.5} \lambda_k \nonumber \\
  &< \epsilon(\lambda_k) + 2d_{\text{max}} L n^{1.5} \lambda_k = \bar \epsilon (\lambda_k).
\end{align}

Since $\lim_{k \to \infty} \lambda_k = 0$, it holds that $\lim_{k \to \infty} \bar \epsilon(\lambda_k) = 0$ and we have that $\lim_{t \to \infty} \bar{\mathcal S}(x(t), p(t)) = 0$. By \eqref{eq:S_convergence}, it holds that $\lim_{t \to \infty} x_{i}(t) \varrho [p_j(t) - p_i(t)]_+ = 0,~ \forall i,j \in \{1, \cdots, n\}$, and using Lemma~\ref{lemma:convergence_revision_protocol}, we conclude that $\lim_{t \to \infty} \inf_{z \in \mathbb{NE}(\mathcal F)} \|x(t) - z\|_2 = 0$. This completes the proof. \hfill \QED

\balance
\bibliographystyle{IEEEtran}
\bibliography{IEEEabrv,references}

\end{document}